\documentclass[twocolumn,prl,showpacs,letterpaper]{revtex4}
\usepackage{graphicx} 
\usepackage{amsmath,amssymb}
\usepackage{dcolumn}

\newcommand{\ud}{\mathrm{d}}
\newcommand{\bx}{\boldsymbol{x}}
\newcommand{\by}{\boldsymbol{y}}

\begin{document} 
\title{Comment on "Irreversibility and Fluctuation Theorem in Stationary Time Series"} 
\author{C. \surname{Van den Broeck}}
\author{B. Cleuren}
\affiliation{Hasselt University - B-3590 Diepenbeek, Belgium}
\begin{abstract}

\end{abstract}
\date{\today}

\maketitle
The characterization of temporal asymmetry is a problem of both fundamental and practical interest. In \cite{porporato}, the authors consider a stationary time series, characterized by the probability $p(x_{1},...,x_{n})$ for the occurrence of the time series $(x_{1},x_{2},...,x_{n})$. Upon introducing the probability distribution $\hat{p}$ for the reversed sequence,
\begin{equation}\label{eq1}
\hat{p}(x_{1},x_{2},...,x_{n})=p(x_{n},x_{n-1}...,x_{1}),
\end{equation}
they propose the relative entropy between $p$ and $\hat{p}$ as a measure of the temporal asymmetry:
\begin{equation}
D(p||\hat{p})=\sum_{\bx} p(\bx)\log\frac{p(\bx)}{\hat{p}(\bx)}.
\end{equation}
with $\bx=(x_{1},...,x_{n})$. The same measure was previously introduced for Markov processes \cite{kurchan,lebowitz} and for dynamical systems \cite{gaspard}, in which context its relation to entropy production was established. On a more microscopic level, a similar expression was recently linked to the dissipation in the case of Hamiltonian dynamics \cite{kawai2007}.\newline Rather than studying ensemble averages, trajectory-dependent quantities obviously carry more detailed information. The authors consider the variable
\begin{equation}
\sigma(\bx)=\log\frac{p(\bx)}{\hat{p}(\bx)},
\end{equation}
which measures the difference in surprise between observing a sequence versus that of observing the time inverse sequence. Further interest in this quantity is stimulated by the observation that its probability distribution $\cal{P}(\sigma)$ of $\sigma$ obeys the so-called fluctuation theorem:
\begin{equation}\label{sigmaft}
\frac{\cal{P}_{\sigma}}{\cal{P}_{-\sigma}}=e^{\sigma}.
\end{equation}
In other words, the probability for a positive value of $\sigma$ is exponentially more likely than that of a negative value. This property is verified for a stochastic process driven by an Ornstein-Uhlenbeck noise and for discharge measurements of the Po river.

In this comment, we point out that the symmetry inherent in the fluctuation theorem is ubiquitous, rendering its physical content less convincing. A similar critique can be opposed to other illustrations of the fluctuation theorem and to the work theorem \cite{crooks}, which has an identical structure. We will indeed show that the fluctuation theorem arises without any restriction for any random variable paired to any  involution \cite{seifertPRL2005,lebowitz,maesPRL2006}.\newline
Let $\bx \in \Omega$ be a set of outcomes with probability distribution $P(\bx)$, and $T$ an involution on the space $\Omega$, namely $T\bx \in \Omega$ and $T^{2}\bx=\bx$. The Jacobian of such a transformation from $\bx$ to $\by=T\bx$ has necessarily an absolute value equal to $1$. Note that time reversal is an example of an involution. In this case, Eq.~(\ref{eq1}) can be written as $\hat{p}(\bx)=p(T\bx)$. Other simple examples correspond to taking the inverse or the negative value of any of the random variables, a suitable rearrangement (such as time reversal) or a combination of all these. \newline
We now define the following random variable:
\begin{equation}
\sigma(\bx)=\log\frac{P(\bx)}{P(T\bx)},
\label{sigma}
\end{equation}
which measures the difference in surprise of observing the outcome $T\bx$ versus that of $\bx$. As a result of the involution, one has that $\sigma(T\bx)=-\sigma(\bx)$. Consider now the probability distribution $\cal{P}_{\sigma}$ for the quantity $\sigma$. We will show that it obeys the fluctuation theorem Eq. (\ref{sigmaft}). The proof follows  directly from the definition Eq.~(\ref{sigma}) and the unit absolute value of the Jacobian (analogue of the Liouville theorem):
\begin{eqnarray}
\cal{P}_{\sigma}&=&\int_{\Omega}\ud \bx \;P(\bx)\;\delta(\sigma(\bx)-\sigma)\\
&=&\int_{\Omega}\ud \bx \;e^{\sigma(\bx)}P(T\bx)\;\delta(\sigma(\bx)-\sigma)\\
&=& e^{\sigma}\int_{\Omega}\ud \by \;P(\by)\;\delta(\sigma(\by)+\sigma)=e^{\sigma}\cal{P}_{-\sigma}.
\end{eqnarray}

\end{document}